# Molybdenum and Nitrogen Co-Doped Titanium Dioxide Nanotube Arrays with Enhanced Visible Light Photocatalytic Activity

Min Zhang*, Juan Wu, Jian Hou, and Jianjun Yang*

*Key Laboratory of Ministry of Education for Special Functional Materials,
Henan University, Kaifeng 475004, P. R. China*

**ABSTRACT**

Molybdenum and nitrogen co-doped $TiO_2$ nanotube arrays (TNAs) were prepared by anodizing in association with hydrothermal treatment. As-prepared Mo and N co-doped TNAs were characterized by field emission scanning electron microscopy, X-ray diffraction, X-ray photoelectron spectroscopy and ultraviolet-visible light diffuse reflection spectroscopy. Besides, the photocatalytic activity of Mo and N co-doped TNAs for the degradation of methylene blue (denoted as MB) under visible light irradiation was evaluated. It was found that N in co-doped TNAs coexists in the forms of N–Ti–O and N–O–Ti, while Mo exists as $Mo^{6+}$ by substituting Ti in the lattice of $TiO_2$. In the meantime, Mo and N co-doping extends the absorption of TNAs into the whole visible light region and results in remarkably enhanced photocatalytic activity for the degradation of MB under visible light irradiation. This could be attributed to the synergetic effect between Mo-doping and N-doping. Namely, Mo and N passivated co-doping produces new states in TNAs, thereby narrowing the bandgap and decreasing the recombination rate of electrons and holes. As a result, the visible light absorption and photocatalytic activity of TNAs is greatly increased. Furthermore, Mo ions with multiple valences in all co-doped samples could act as trapping sites to effectively decrease the recombination rate of electrons and holes, which also contribute to improved photocatalytic activity of Mo and N co-doped TNAs.

## 1. INTRODUCTION

Highly ordered $TiO_2$ nanotube arrays (TNAs) prepared by anodizing technique have received considerable attention in recent years, due to their large surface area, high photocatalytic activity, and good recyclability as well as unique nanostructure and promising application in solar light conversion as compared with conventional $TiO_2$ powder.[1,2] Unfortunately, the use of $TiO_2$ as a visible light ($\lambda = 400 \sim 750$ nm, constituting $\sim 45\%$ of the solar spectrum) photocatalyst is greatly limited, since it has a large bandgap of 3.2 eV and can only be activated by ultraviolet (denoted as UV) radiation ($\lambda < 387$ nm) that constitutes only a small fraction ($3\% \sim 5\%$) of the solar spectrum.[3]

To overcome the above-mentioned drawback, many researchers have made great efforts to modify the band structure of $TiO_2$ so as to shift its absorption edge towards the visible light region.[4,5] For this purpose, mono-doping of $TiO_2$ with $3d$ transitional metals or nonmetals has been extensively explored. For example, Yu et al. synthesized nitrogen self-doped $TiO_2$ nanosheets with exposed {001} facets by solvothermal treatment of TiN in an ethanol solution of $HNO_3$–HF and acquired much higher visible light photocatalytic $H_2$-production activity as compared with nitrogen doped $TiO_2$ microcrystallites.[6] Yang et al. prepared carbon-doped anatase $TiO_2$ by mild oxidation of TiC and found that C-doped $TiO_2$ powders exhibited photoactivity to trichloroacetic acid degradation under visible light irradiation.[7] However, mono-doped $TiO_2$ systems usually contain partially occupied impurity bands that can act as recombination centers to counteract the doping efficiency thereby adding to difficulty in control of proper synthesis process and parameters.[8,9] Besides, although anodization in the presence of electrolyte containing $NH_4F$ can be well adopted to incorporate nitrogen within the nanotubes of TNAs that grow from the metal by electrochemical oxidation, this traditional solution-based N doping technique has a very low N doping level, and its bandgap narrowing effectiveness is often compromised by numerous recombination centers causing loss of photogenerated electron–hole pairs.[1,4,10] This is why co-doping of $TiO_2$ by $3d$

*Authors to whom correspondence should be addressed.
Emails: zm1012@henu.edu.cn, yangjianjun@henu.edu.cn

transitional metals (such as Mo, W and V) and nonmetals is of particular significance.[11, 12] To name a few, Yu et al. using a simple mixing-calcination method, prepared nitrogen and sulfur co-doped $TiO_2$ nanosheets with exposed {001} facets and found that as-prepared N–S–$TiO_2$ samples exhibit enhanced visible-light photocatalytic activity, due to the intense absorption in the visible light region and the exposure of highly reactive {001} facets of $TiO_2$ nanosheets.[13] Kubacka et al. synthesized W and N co-doped $TiO_2$ anatase $TiO_2$ possessing both unprecedented high activity and selectivity in the gas-phase partial oxidation of aromatic hydrocarbons under sunlight irradiation in the presence of molecular oxygen as the oxidant.[14] Tan et al. prepared $TiO_2$ nanocrystalline co-doped with molybdenum and nitrogen by a sol–gel method and found that the co-doped sample exhibits excellent visible-light absorption performance.[15] Liu et al. prepared a series of $Ti_{1-x}Mo_xO_{2-y}N_y$ samples using sol–gel method and found that Mo and N co-doping (in particular, passivated co-doping) contributes to increase the up-limit of dopant concentration and create more impurity bands in the bandgap of $TiO_2$ thereby considerably increasing visible light photocatalytic activity.[16] Yin et al. and Long et al. based on density-functional theory calculations, separately proposed that the injection of $3d$ or $4d$ transition metals (such as Mo, W and V) can increase the solubility limits of N and the band edges of $TiO_2$ can be modified by co-dopants to significantly shift the valence band edge up.[17, 18]

Nevertheless, few reports are currently available about Mo and N co-doping of TNAs and their photocatalytic activity as well. Therefore, in this research we adopt a sequence of anodization technique and hydrothermal synthesis technique to fabricate Mo and N co-doped TNAs. This article reports the morphology, structure and optical absorption performance of as-fabricated Mo and N co-doped TNAs as well as the influence of doping concentration on their photocatalytic activity for the degradation of methylene blue (denoted as MB) under visible light irradiation ($\lambda > 420$ nm).

## 2. EXPERIMENTAL DETAILS

### 2.1. Synthesis of Mo and N Co-Doped $TiO_2$ Nanotube Arrays

Self-organized and well-aligned TNAs were fabricated by two-step electrochemical anodization process.[19] Briefly, titanium sheets (20 mm × 40 mm, purity > 99.6%; purchased from Baoji Boxin Metal Materials Company Ltd. (Baoji, China)) with a thickness of 0.25 mm were sequentially sonicated in acetone, isopropanol and methanol for 10 min, followed by etching in the mixture of $HF/HNO_3/H_2O$ (1:4:5 in volume) for 20 s, rinsing with deionized water and drying under a $N_2$ stream. Resultant rectangular Ti sheet was used as an anode to couple with Pt meshwork as a cathode in the anodic oxidation test set-up. A direct current power supply and a mixed electrolyte solution of ethylene glycol containing 0.25 wt% $NH_4F$ and 2 vol% deionized water were used in the electrochemical processes. The first oxidation step was conducted at 60 V for 1 h, with which the oxidized surface films were removed by sonication in distilled water before the oxidized samples were dried under high purity $N_2$ stream at room temperature. As-dried Ti substrates were then oxidized in the original electrolyte at 60 V for 2 h, followed by sonication in ethanol for 8 min to remove the surface cover and drying under high purity $N_2$ stream to afford as-prepared TNAs. Magnetic stirring was adopted throughout the oxidation processes while a circulation pump was performed at a low temperature of about 20 °C.

Before Mo and N co-doping, as-prepared TNAs were calcinated at 500 °C for 3 h in a furnace (heating rate 10 °C/min), corresponding sample was denoted as Mo–N–$TiO_2$–0. Interstitial nitrogen species were formed in Mo–N–$TiO_2$–0 due to the electrolyte containing $NH_4F$ or a lesser extent atmospheric nitrogen dissolved into the electrolyte during anodization.[10] Mo and N co-doped TNAs with different dopant concentration (0.05%, 0.10% and 1.00%; referring to different concentration of Mo and N source) were then prepared by hydrothermally treating Mo–N–$TiO_2$–0 in a Teflon-lined autoclave (120 mL, Parr Instrument) containing approximately 60 mL of ammonium molybdate solution (($NH_4)_6Mo_7O_{24} \cdot 4H_2O$) as the source of both Mo and N. The hydrothermal synthesis was conducted at 180 °C for 5 h in a box oven, and then the autoclave was naturally cooled down to room temperature affording target products denoted as Mo–N–$TiO_2$–0.05, Mo–N–$TiO_2$–0.1, Mo–N–$TiO_2$–1 (numeral suffixes 0.05, 0.1 and 1 refer to ammonium molybdate solution concentration of 0.05%, 0.10% and 1.00%) after rinsing with distilled water and drying under high purity $N_2$ stream at room temperature.

### 2.2. Characterization

Surface morphology of TNAs was observed using a field emission scanning electron microscope (FESEM). A Philips X'Pert Pro PW3040/60 X-ray diffractometer (XRD, Philips Corporation, Holland) was performed to determine the crystal structure of as-prepared samples. Chemical state analysis was conducted with an Axis Ultra X-ray photoelectron spectroscope (XPS, Kratos, UK; a monochromatic Al source operating at 210 W with a pass energy of 20 eV and a step of 0.1 eV was used). All XPS spectra were corrected using the C $1s$ line at 284.8 eV. A Cary-5000 scanning ultraviolet-visible light diffuse reflection spectrophotometer (denoted as UV-vis DRS; Varian, USA) equipped with a Labsphere diffuse reflectance accessory was performed to collect the UV-vis diffuse reflectance spectra. The adsorption spectra of undoped and Mo + N co-doped TNAs in the range of 300∼700 nm were also recorded with the UV-vis DRS facility at a scan rate of 600 nm min$^{-1}$.

### 2.3. Photocatalytic Activity Measurements

Photocatalytic activities of all samples were evaluated by monitoring the photocatalytic degradation of MB in aqueous solution under visible light with vertical irradiation as light source. Briefly, to-be-tested Mo–N–TiO$_2$–X (X = 0, 0.05, 0.1 and 1) samples were dipped into 25 mL of MB solution with an initial concentration of 10 mg L$^{-1}$. The effective photocatalytic reaction area of TNAs films was 4 cm$^2$. Prior to irradiation, the MB solution containing to-be-tested sample was magnetically stirred in the dark for 30 min to establish adsorption–desorption equilibrium. During the photocatalytic reaction, the absorbance of MB at 664 nm was measured using an SP-2000 spectrophotometer at a time interval of 30 min. The decoloration rate of MB solution is calculated as $(C_0 - C)/C_0 \times 100\%$, where $C_0$ is the concentration of MB at adsorption–desorption equilibrium in the dark and $C$ is the concentration of MB upon completion of the photocatalytic reaction under visible light irradiation. In this way, the influence of adsorption amount of MB on its solution decoloration rate is excluded (different photocatalyst samples adsorb different amount of MB).

## 3. RESULTS AND DISCUSSION

### 3.1. SEM Images

Top-view SEM images of as-prepared Mo–N–TiO$_2$–X catalysts are presented in Figure 1. The undoped nanotubes of TNAs are open at the top end and have an average diameter of approximately 100 nm (Fig. 1(a)), and their nanotubular array structure is clearly observed after the surface layers are slightly removed with a razor blade (see the inset of Fig. 1(a)). Besides, TNAs co-doped with a small concentration of Mo and N have morphology similar as that of undoped counterpart (comparing Figs. 1(b) and (c) with Fig. 1(a)). This reveals that co-doping low concentration of Mo and N does not destroy the tubular structure of TNAs, although a small amount of particles appears on the surface of Mo–N–TiO$_2$–0.1 nanotube arrays (see the circles in inset of Fig. 1(c)). However, co-doping high concentration (1.00%) of Mo and N leads to more particles and destroys the nanotubular structure (Fig. 1(d)), which is possibly related to the formation of MoO$_3$ particles.[15]

### 3.2. XRD Analysis

The XRD patterns of various samples are shown in Figure 2. It can be seen that all the diffraction peaks of undoped TNAs and various co-doped TNAs can be ascribed to anatase TiO$_2$, which indicates that Mo and N co-doping has no effect on the crystal structure and phase composition of TNAs. Besides, although SEM observations confirms that there exists a small amount of particles on the surface of co-doped TNAs, no significant characteristic peak of MoO$_3$ is found in corresponding XRD patterns, which could be attributed to low dopant concentration or even scattering of MoO$_3$. Moreover, Mo and N co-doping causes the diffraction peak of anatase TiO$_2$ along (101) plane to shift towards higher values of $2\theta$. This implies that the O or Ti atoms in the anatase lattice of co-doped samples may be substituted by N and Mo atoms, since Mo(VI) ions with a radius (0.062 nm) close to that of Ti$^{4+}$ (0.068 nm) are easy to enter into the lattices of TiO$_2$ and displace Ti$^{4+}$ ions while nitrogen and molybdenum have a synergistic doping effect.[20, 21]

### 3.3. XPS Analysis

To investigate the chemical states and the concentration of N and Mo atoms incorporated into Mo and N co-doped TNAs, we conducted XPS analysis. Figure 3(a) shows the high resolution XPS spectra and corresponding fitted XPS spectra for the N 1s region of Mo–N–TiO$_2$–X nanotube arrays. All co-doped Mo–N–TiO$_2$–X samples show a broad N 1s peak within 396∼403 eV. However, undoped Mo–N–TiO$_2$–0 sample shows a N 1s peak at 399.6 eV, and deconvolution of N 1s peak reveals the presence of N with a binding energy of 398 eV and 399.8 eV for Mo and N co-doped TNAs, respectively. To date, the assignment of the N 1s peak of doped TiO$_2$ is still under debate. For example, Asahi et al. reported that the N 1s peak of N-doped TiO$_2$ at 396 eV is attributed to atomic $\beta$-N state which acts as the vis-photocatalytic active site.[5] Nevertheless, many recent studies claimed the absence of the N 1s peak at about 396 eV and the presence of a broad N 1s peak in a range of 397∼403 eV.[22–26] Particularly, Cong et al. attributed the broad N 1s peak within 396∼403 eV to O–Ti–N (substitutional N) and Ti–O–N (interstitial N) structures.[26] Considering this viewpoint, we suggest that the N 1s peak at 398 eV can be ascribed to anionic N incorporated in TiO$_2$ via O–Ti–N linkage, while the N 1s peak at 399.8 eV can be attributed to oxidized nitrogen like N–O–Ti.[3, 5, 27]

Undoped TNAs and TNAs co-doped with Mo and N show N 1s peaks with obvious difference, which implies that Mo and N co-doping leads to changes in the chemical state of N species. The single N 1s peak of undoped TNAs at 399.6 eV is assigned to interstitial nitrogen species, as reported by Grimes et al. who suggested that the nitrogen incorporated within TiO$_2$ nanotubes during anodization is primarily supplied by NH$_4$F or low extent of atmospheric nitrogen dissolved into the electrolyte.[16] Besides, the broad N 1s peak of various Mo and N co-doped samples within 396∼403 eV gives evidences to the coexistence of substitutional and interstitial nitrogen species.

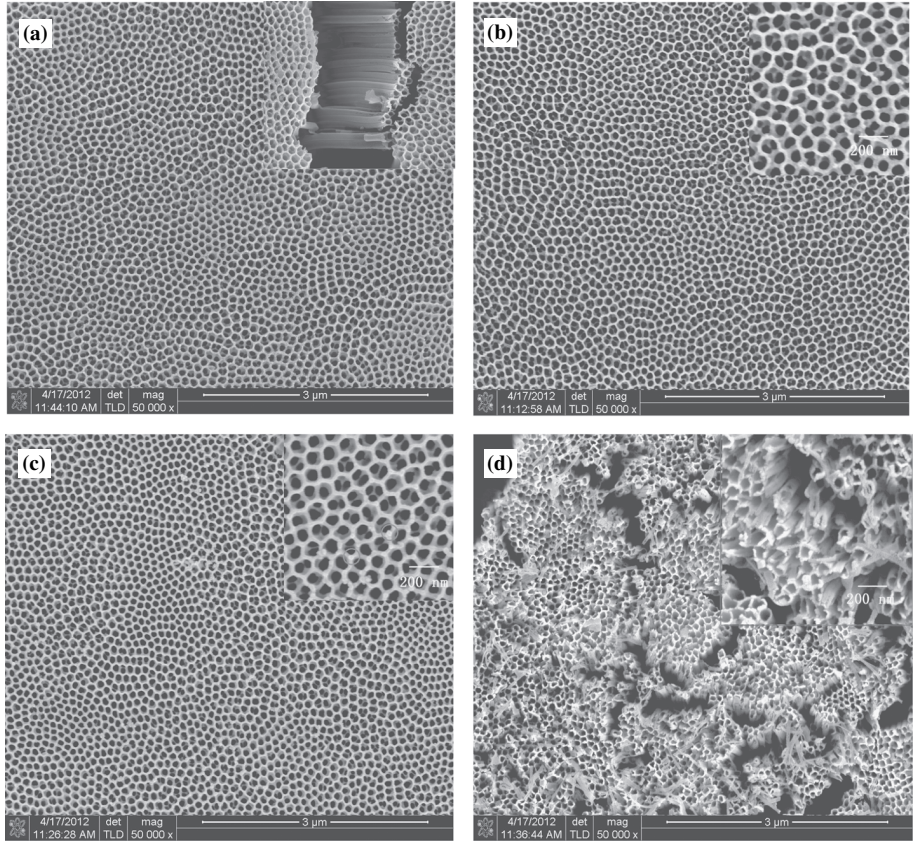

**Fig. 1.** FESEM cross-sectional views of Mo and N co-doped $TiO_2$ nanotube arrays: (a) Mo–N–$TiO_2$–0, (b) Mo–N–$TiO_2$–0.05, (c) Mo–N–$TiO_2$–0.1, and (d) Mo–N–$TiO_2$–1 (inset shows the magnified view).

High resolution Mo $3d$ spectra of Mo and N co-doped TNAs are presented in Figure 3(b). The Mo $3d$ peaks at about 232.2 eV and 235.5 eV are assigned to $3d_{3/2}$ and $3d_{5/2}$ electronic states of $Mo^{6+}$, respectively.[28] This indicates that Mo substitutes Ti in the lattice of $TiO_2$ and exists in an oxidation state of $Mo^{6+}$, which is accordance with relevant XRD analytical results.

The relative atomic concentration (abridged as at.%) of sample Mo–N–$TiO_2$–0 and TNAs co-doped with Mo and N, determined by quantitative XPS analysis accounting

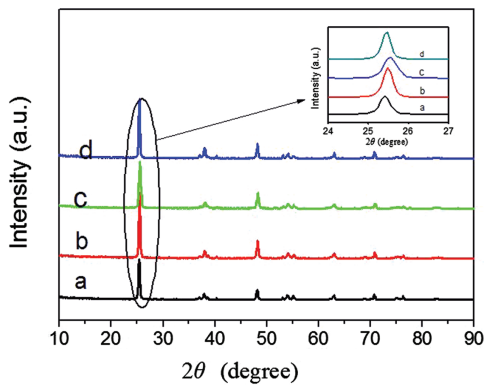

**Fig. 2.** XRD patterns of (a) Mo–N–$TiO_2$–0, (b) Mo–N–$TiO_2$–0.05, (c) Mo–N–$TiO_2$–0.1, and (d) Mo–N–$TiO_2$–1.

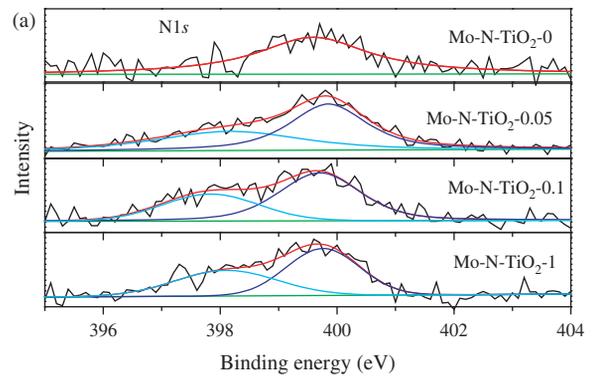

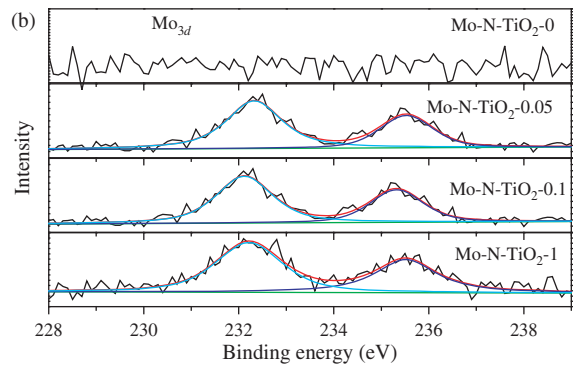

**Fig. 3.** Curve-fitted of N $1s$ spectrum (a) and Mo $3d$ spectrum (b) of Mo–N–$TiO_2$–X.

Table I. Surface atomic concentration of Mo–N–TiO$_2$–X samples determined by quantitative XPS analysis.

| Samples | Atomic concentration (at.%) | | | |
| --- | --- | --- | --- | --- |
| | Mo[a] | N[b] | Ti[c] | O[d] |
| Mo–N–TiO$_2$–0 | 0 | 0.63 | 33.90 | 65.48 |
| Mo–N–TiO$_2$–0.05 | 0.14 | 2.33 | 29.18 | 68.35 |
| Mo–N–TiO$_2$–0.1 | 0.36 | 3.57 | 30.27 | 65.79 |
| Mo–N–TiO$_2$–1 | 0.44 | 4.70 | 29.78 | 65.09 |

*Note*: [a,b,c,d] Calculated according to the peak areas of Mo 3$d$, N 1$s$, Ti 2$p$ and O 1$s$, respectively.

for the peak areas of Mo 3$d$, N 1$s$, Ti 2$p$ and O 1$s$, is presented in Table I. Before co-doping, sample Mo–N–TiO$_2$–0 has a N doping concentration of 0.63 at.%. Besides, the atomic concentration of doped Mo increases significantly with the increase of ammonium molybdate solution concentration, and that of doped nitrogen increases simultaneously upon doping of Mo into the lattice of TiO$_2$. This demonstrates that Mo and N co-doping is in favor of enhancing the dopant solubility. Particularly, the Mo and N doping levels of sample Mo–N–TiO$_2$–1 prepared at the maximum ammonium molybdate solution concentration of 1.00% are the highest, and they are also much larger than that of sample Mo–N–TiO$_2$–0. This is possibly because Mo and N co-doping has synergistic effect and leads to a charge compensation among donors and acceptors thereby greatly enhancing the solubility limit of N in TiO$_2$.[29, 30]

### 3.4. UV-Vis Diffuse Reflectance Spectrometric Analysis

The UV-vis diffuse reflectance spectra of various samples are presented in Figure 4. As reported elsewhere, TiO$_2$ nanotube arrays show a broad absorption around 400∼500 nm which is probably caused by the trapped charge carriers or the absorption of incident light by nanotube arrays.[31, 32] In the present research, sample Mo–N–TiO$_2$–0 has relatively weak visible-light response, which is attributed to interstitial N dopant (as evidenced by N 1$s$ peak at about 399.6 eV) that induces the local states near the valence band edge and narrows the bandgap. Different from sample Mo–N–TiO$_2$–0, Mo and N co-doped TNAs have red-shifted absorption edge and strong absorption in the visible-light region, and their absorption tail is extended to about 570 nm. Particularly, the visible-light absorption of Mo and N co-doped TNAs is enhanced with increasing dopant concentration. This is because Mo and N incorporated in the lattice of TNAs have synergistic doping effect which favors to increase the solubility limits of Mo and N thereby greatly enhancing the visible-light absorption along with the increase of the up-limits of Mo and N concentration.[29, 30] Specifically, a larger dopant concentration refers to a higher impurity level which is favorable for further increasing the visible-light absorption of TiO$_2$, thus sample Mo–N–TiO$_2$–1 exhibits the highest intensity of visible light adsorption.

### 3.5. Evaluation of Photocatalytic Activity

The photocatalytic activity of Mo–N–TiO$_2$–X samples was evaluated by monitoring the degradation of MB under visible light irradiation. As shown in Figure 5, Mo and N co-doped TNAs possess much higher photocatalytic activity under visible light irradiation than sample Mo–N–TiO$_2$–0. Besides, the decoloration rate of MB under visible light irradiation in the presence of Mo and N co-doped TNAs increases with the increase of dopant concentration; and in particular, sample Mo–N–TiO$_2$–1 exhibits the best photocatalytic efficiency for the degradation of MB and allows nearly complete elimination of MB within 3 h.

The origin of the visible light photocatalytic activity of Mo and N co-doped TNAs can be attributed to the

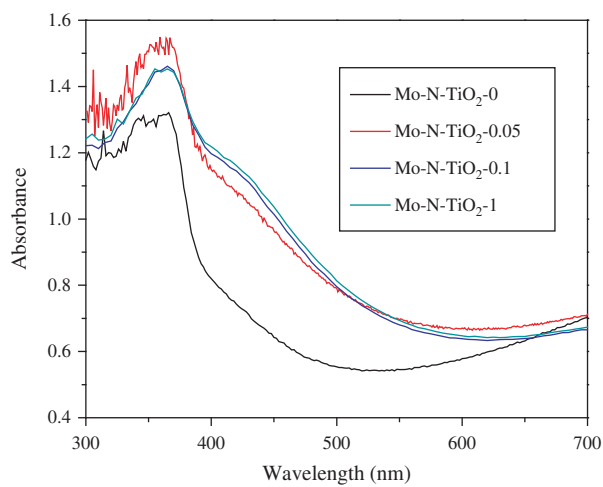

**Fig. 4.** UV-vis diffuse reflectance spectra of Mo–N–TiO$_2$–X.

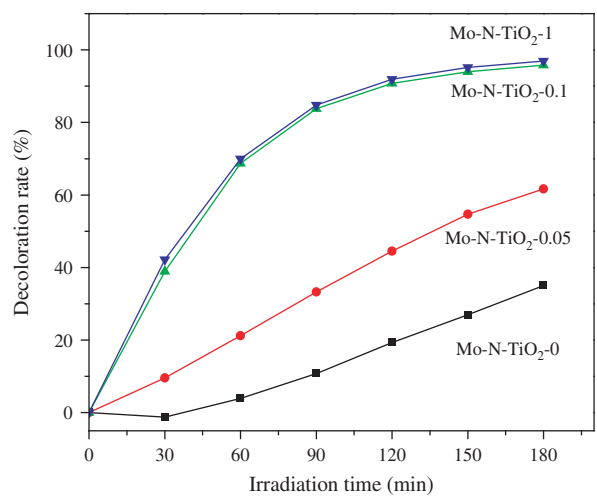

**Fig. 5.** Photocatalytic degradation rate of MB under visible light irradiation in the presence of Mo–N–TiO$_2$–X photocatalysts.

interactions between N-doping and Mo-doping. Before co-doping, interstitial N doping can induce the local states near the valence band edge and narrow the bandgap, which results in the visible-light photocatalytic activity of sample Mo–N–TiO$_2$–0. After co-doping with Mo and N, Mo substitutes for Ti in the lattice and exists in the form of Mo$^{6+}$. The doping energy level of Mo$^{6+}$/Mo$^{5+}$ is 0.4 eV, and it is much more positive than the potential of the conduction band of TiO$_2$ particles [$E_{cb} = -0.5$ eV vs. NHE (normal hydrogen electrode) at pH = 1].[31, 33] As a result, after Mo is incorporated into the lattice of TiO$_2$, substitutional N atoms and interstitial N dopant coexist in the lattice, while Mo dopant allows more N atoms to be incorporated into the lattice of TiO$_2$. In this way, while N is doped into the substitutive sites of TiO$_2$ during Mo and N co-doping, the N 2$p$ accepter state functions to narrow the bandgap of TiO$_2$ via mixing with O 2$p$ states.[34] Consequently, the bandgap energy of TiO$_2$ is significantly lowered and its absorption in the visible-light region is considerably increased, which makes it feasible for Mo and N co-doped TNAs to be activated by visible light while more electrons and holes are generated to participate in the photocatalytic reactions. Moreover, the dopant concentration plays a very important in increasing the photocatalytic activity of Mo and N co-doped TNAs in visible light region. This is because a small amount of Mo$^{6+}$ can act as a temporary photo-generated electron or hole-trapping site thereby inhibiting the recombination of photo-generated charge carriers and prolonging their lifetime.[35, 36] The detailed reaction steps are as follows:

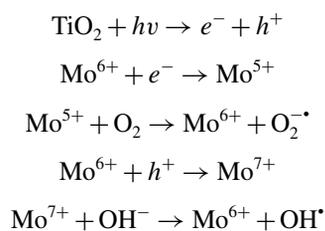

$$TiO_2 + h\nu \rightarrow e^- + h^+$$
$$Mo^{6+} + e^- \rightarrow Mo^{5+}$$
$$Mo^{5+} + O_2 \rightarrow Mo^{6+} + O_2^{-\bullet}$$
$$Mo^{6+} + h^+ \rightarrow Mo^{7+}$$
$$Mo^{7+} + OH^- \rightarrow Mo^{6+} + OH^{\bullet}$$

When Mo$^{6+}$ ($d^5$) traps an electron, its electronic configuration is transferred to $d^6$; and if it traps a hole, its electronic configuration will be transferred to $d^4$ which is highly unstable. Therefore, to restore the stable electronic configuration, the trapped charge carrier tends to be transferred from Mo$^{5+}$ or Mo$^{7+}$ to the adsorbed O$_2$ or surface hydroxyl (OH$^-$) thereby regenerating Mo$^{6+}$. These newly produced active species (such as OH$^{\bullet}$ and O$_2^{-\bullet}$) are able to initiate the photocatalytic reactions. Though the Mo$^{6+}$/Mo$^{5+}$ energy level can be the recombination center if the dopant concentration is not proper, Mo and N co-doped TiO$_2$ may have a synergistic doping effect of Mo dopant and N dopant to inhibit the recombination of photogenerated holes and electrons at the Mo$^{6+}$/Mo$^{5+}$ doping energy level. Namely, Mo and N co-doping leads to charge compensation among donors and acceptors. Since the charges of the donors and acceptors are fully compensated during passivated doping, the impurity bands are passivated to be less effective recombination centers owing to the proposed equilibrium charge mechanism. As a result, the number of carrier recombination centers is reduced and the photocatalytic activity is enhanced.[15] Therefore, the best visible light photocatalytic activity of sample Mo–N–TiO$_2$–1 with the highest dopant concentration among various tested photocatalysts can be attributed to its much larger number of impurity states than samples Mo–N–TiO$_2$–0.05 and Mo–N–TiO$_2$–0.1. It is Mo and N ions incorporated into the lattice of TiO$_2$ that alter the crystal and electronic structure of TNAs and result in significantly improved photocatalytic activity in the visible light region.

## 4. CONCLUSIONS

Mo and N co-doped TiO$_2$ nanotube arrays have been successfully synthesized by combining anodizing technique with hydrothermal method. XPS data reveal that, in as-prepared Mo and N co-doped TNAs, N coexists in the forms of N–Ti–O and N–O–Ti corresponding to substitutional N atoms and interstitial N dopant, while Mo substitutes Ti in the lattice of TiO$_2$ and exists in the oxidation state of Mo$^{6+}$. UV-vis DRS data show that Mo and N co-doping extends the absorption of TiO$_2$ nanotube arrays into the whole visible light region. There exists synergetic doping effect between Mo and N, which plays a key role in producing new states, narrowing the bandgap and reducing the recombination effectively thereby greatly improving the visible light absorption and photocatalytic activity of TNAs. Besides, Mo ions with multiple valences in Mo–N–TiO$_2$–X samples can act as trapping sites to effectively decrease the recombination rate of electrons and holes, also resulting in improved photocatalytic activity of TNAs. Therefore, Mo and N co-doped TiO$_2$ nanotube arrays, in particular, sample Mo–N–TiO$_2$–1 with the highest visible light absorption and photocatalytic activity, could be promising visible light photocatalysts.

**Acknowledgments:** The authors thank the National Natural Science Foundation of China (grant Nos. 20973054 and 21203054) and the Science and Technology Department of Henan Province (natural science project grant No. 112300410171) for financial support.